\documentclass[12pt,showpacs,preprintnumbers,amsmath,amssymb]{revtex4}
\usepackage{graphicx}
\usepackage{dcolumn}
\usepackage{bm}

\newcommand{\bi}{\begin{itemize}}
\newcommand{\ei}{\end{itemize}}
\newcommand{\be}{\begin{equation}}
\newcommand{\ee}{\end{equation}}
\newcommand{\ba}{\begin{eqnarray}}
\newcommand{\ea}{\end{eqnarray}}
\newcommand{\bse}{\begin{subequations}}
\newcommand{\ese}{\end{subequations}}
\newcommand{\M}{{\cal {M}}}

\newcommand{\CZ}{{\cal {Z}}}

\newcommand{\rvir}{r_{\textrm{\tiny{vir}}}}
\newcommand{\Mvir}{M_{\textrm{\tiny{vir}}}}

\newcommand{\dd}{\textrm{d}}

\begin{document}


\title{Stellar
polytropes and Navarro--Frenk--White dark matter halos: a connection to
Tsallis entropy.}

\author{Luis G. Cabral-Rosetti$^\dagger$, Tonatiuh Matos$^\ddagger$,
Dar\'\i o N\'u\~nez$^\dagger$, Roberto A Sussman$^\dagger$,
Jes\'us Zavala$^\dagger$}


\affiliation{$^\dagger$Instituto de Ciencias Nucleares
Universidad Nacional Aut\'onoma de M\'exico \\
A. P. 70-543,  M\'exico 04510 D.F., M\'exico. \\
$^\ddagger$ Departamento de F{\'\i}sica,
Centro de Investigaci\'on y de Estudios Avanzados del IPN, \\
A.P. 14-740, 07000 M\'exico D.F., M\'exico.}


\begin{abstract}
We present an alternative for the description of galactic
halos based on Tsallis' non--extensive entropy formalism; on this scheme, halos are
stellar polytropes characterized by three parameters, the central density, $\rho_c$, the
central velocity dispersion, $\sigma_c$ and the polytropic index, $n$. To evaluate these parameters
we take the Navarro-Frenk-White paradigm as a comparative model and make the following assumptions: both
halo models must  have the same virial mass, the same total energy and the same maximal velocity.
These three conditions fix all the parameters for a given stellar polytrope allowing us to compare
both halo models. The halos studied have virial masses on the range $10^{12}-10^{15}\,M_\odot$, and it
was found after the analysis that they are described, at all scales, by almost the same
polytropic index, $n\approx 4.8$, implying an
empirical estimation of Tsallis non--extensive parameter for this type of dynamical
systems: $q\approx
1.3$.
\end{abstract}


\maketitle
\section{Introduction}

Cold Dark Matter (CDM) models based on N--body numerical
simulations predict excessive substructure and cuspy dark matter
halo profiles that are not observed in the rotation curves of
dwarf and LSB galaxies \cite{Blok1}. The significance of this
discrepancy with observations is still under dispute, leading to
various theoretical alternatives, either within the thermal
paradigm (self--interacting \cite{self} and/or ``warm''
\cite{warm} DM, made of lighter particles), or non--thermal dark
matter (DM) models (real \cite{DMCQG} or complex \cite{Ruffini}
scalar fields, axions, etc). However, none of these alternatives
is free of controversy. On the other hand, the CDM model of
collision--less WIMP's could remains still as a viable model to
account for DM in galactic halos, provided there is a mechanism to
explain the discrepancies of this model with observations in the
center of galaxies. The main goal of this paper is to give an
approach for such a possibility. The main idea for this
alternative is as follows. Since gravity is a long--range
interaction and virialized self--gravitating systems are
characterized by non--extensive forms of entropy and energy, it is
reasonable to expect that the final configurations of halo
structure predicted by N--body simulations must be, somehow,
related with states of relaxation associated with non--extensive
formulations of Statistical Mechanics. 

The usual statistical mechanic treatment of self--gravitational
systems is provided by the micro-canonical ensemble, which is
compatible with negative heat capacities associated with known
effects, such as gravothermal instability~\cite{Padma2,Padma3}.
An alternative formalism that allows non-extensive forms for
entropy and energy has been developed by Tsallis~\citep{Tsallis}
and applied to self--gravitating systems~\citep{PL,TS1,TS2}, under
the assumption of a kinetic theory treatment and a mean field
approximation. As opposed to the Maxwell--Boltzmann distribution
that follows as the equilibrium state associated with the  usual
Boltzmann--Gibbs entropy functional, the Tsallis functional
yields  as equilibrium state the ``stellar polytrope'',
characterized by a polytropic index $n$. The stellar polytrope
yields a Maxwell--Boltzmann distribution function (the isothermal sphere) in the
limit $n\to\infty$. This index is related to the
``non--extensivity'' parameter $q$ of Tsallis entropy functional,
so that the ``extensivity'' limit $q\to 1$ corresponds to the
isothermal sphere.

Although the self--gravitating  collision-less and virialized gas
that makes up galactic halos is far from the state of
gravothermal catastrophe, it is reasonable to assume that it is
near some form of relaxation equilibrium characterized by
non--extensive forms of entropy and energy. On the other hand,
high precision N--body  numerical simulations, perhaps the most
powerful method available for understanding gravitational
clustering, roughly yield density, mass and rotation velocity
profiles that seem to fit observed galactic halo structures
(pending the controversy on excess substructure and cuspy density
cores specially on LSB galaxies). Admitting that stellar polytropes follow from an idealized
approach based on kinetic theory and an isotropic distribution
function, it is still interesting to verify empirically if the
structural parameters of the halo gas, or at least of the outcome
of numerical simulations, can be adjusted to those of a stellar
polytrope, the equilibrium state under Tsallis' formalism. 

It is important to notice that the main objective of this paper is 
not to compare this polytropic model of dark halo with observational 
results coming from actual galaxies but with the Navarro-Frenk-White 
paradigm of dark halos \cite{NFW} which adequately describes the rotation curves of
most galaxies. It is known
that the NFW profile fits well the density profile of galaxies in
the region outside their core. It fails in the
central regions where observations show that the
density profile is almost flat. In this work we show that, for
example, the best polytropic fit to halos with NFW profiles
follows from polytropes characterized by densities in the range
$3.7\times10^{-4}\,\text{M}_\odot/ \text{pc}^3\,<\rho_c\,<\,
1.2\times10^{-3}\,\text{M}_\odot/\text{pc}^3$ and polytropic
indices is almost constant with a value near $n=4.8$, or $q\approx1.3$

Since these best--fit polytropes have the same observable
quantities as the NFW halos without central density cusps, they
might provide an even better fit to halo structures than the usual
NFW profiles. Furthermore, the present analysis and results can be
used to calibrate the values of the free parameter $q$ that
emerges from Tsallis's formalism.

Hence, we propose to verify which parameters of the stellar polytropes 
provide a suitable description of the halo that resembles the one
that emerges from the well known numerical simulations of Navarro--Frenk--White.
For the wide virial mass range of $10^{10}<\Mvir/M_\odot
<10^{15}$.

The paper is organized as follows: in section II we provide the equilibrium
equations of stellar polytropes and briefly summarize their connection
to Tsallis' non--extensive entropy formalism. In section III we discuss the
dynamical variables of NFW halos, while in section IV 
we describe a procedure to compare a polytropic halo with an NFW one and obtain numerically
the parameters that characterize such polytropic halos, producing graphics showing such
comparison. A summary of our results is given in
section V.

\section{Tsallis entropy and stellar polytropes.}

For a face space given by $({\bf{r}},\,{\bf{p}})$, the kinetic
theory entropy functional associated with Tsallis' formalism
is \cite{PL,TS1}, and \cite{TS2}
\begin{equation}
S_{q}\ =\ -\frac{1}{q-1}\,\int
{(f^{q}-f)\,{d}^{3}{\bf{r}}\,%
{d}^{3}{\bf{p}}},  \label{q_entropy}
\end{equation}
where $f$ is the distribution function and $q>1$ is a real number. In
the
limit $q\rightarrow 1$, the functional (\ref{q_entropy}) leads to the
usual
Boltzmann--Gibbs functional. The condition $\delta \,S_{q}=0$ leads to the
distribution function that corresponds to a stellar polytrope characterized by
the equation of state
\begin{equation}
p\ =\ K_n\,\rho ^{1+1/n}, \label{eq:edo}
\end{equation}
where $K_n$ is a function of the polytropic index $n$, and can be expressed
in terms of the central parameters:
\begin{equation} K_n=\frac{{\sigma_c}^2}{{\rho_c}^{1/n}}.
\label{eq:Kn}\end{equation}
The polytropic index, $n$, is related to the Tsallis' parameter $q>1$ by:
\begin{equation}
n\ =\ \frac{3}{2}+\frac{1}{q-1}\end{equation}
The stability condition $\delta^{2}S_{q}<0$ is only satisfied generically
for polytropes with $3/2 <n<5$ (or $9/7<q$), which are then stable equilibrium
configurations. Polytropes with $n>5$ are then meta--stable configuration which
undergo gravothermal instability for sufficiently large density contrast
$\rho/\rho_c$ (see \cite{TS1,TS2}).

The standard approach for studying spherically symmetric hydrostatic
equilibrium in stellar polytropes follows from inserting (\ref{eq:edo}) into
Poisson's equation, leading to the well known Lane--Emden equation~\cite{B-T}
\begin{eqnarray}
\frac{1}{x^2}\,\frac{{d}}{{d}
x}\left(x^2\,\frac{{d}\,\theta}{%
{d} x}\right)+\theta^n \ = \ 0,  \label{eq:LE}
\end{eqnarray}
with
\ba \theta \ &=& \ \left(\frac{\rho}{\rho_c}\right)^{1/n}
\\ x \ &=& \ \frac{r}{r_0},\quad r_0^{-2} \ = \
\frac{4\pi G\,
\rho_c}{\sigma_c^2
}, \quad \sigma_c^2 \ = \ \frac{p_c}{\rho_c},\\
&G \ &= \ 4.297\times 10^{-6}\,\frac{\left(\rm{km/sec}\right)^2}
{\rm{M}_\odot/\rm{kpc}}
\ea
where $\sigma_c$ and $\rho_c$ are the central velocity dispersion
and central mass density respectivelly; and we take this value for the gravitational constant due to
the units we are using. Notice that the velocity dispersion is a
measure of the kinetic temperature of the gas by the relation:
$\sigma_c^2=k_{_B}\,T_c/m$, with $k_{_B}$ being Boltzmann's
constant, and that we are using a normalization for $r_0$, which
differs from the usual one by a factor $1/(n+1)$. We find it more
convenient to consider instead of (\ref{eq:LE}) the following set
of equivalent equilibrium equations
\begin{eqnarray}
\frac{\dd \M}{\dd x} && = \ x^2\,\CZ,
\nonumber \\
\frac{\dd\CZ}{\dd x} && = \
-\frac{n}{n+1}\,\frac{\M\,\CZ^{1-1/n}}{x^2}, \label{eq_ZMx}
\end{eqnarray}
where $\M$ and $\CZ$ relate to $M,\,\rho$
(mass and mass density at radius $r$) by
\begin{eqnarray}
\M && = \ \frac{M}{4\pi \rho_c\,r_0^3}, \qquad \CZ
\ = \ \frac{\rho}{\rho_c},
  \label{pars}
\end{eqnarray}
Notice that in the limit $n\to\infty $, equations
(\ref{eq_ZMx}) become the equilibrium equations of the isothermal sphere.

Once the system (\ref{eq_ZMx}) has been integrated numerically,
the velocity profile derived from the virial theorem takes the form:
\begin{equation}V^2(x) \ = \ \sigma_c^2\,\frac{\M}{x},
\label{rotvel}\end{equation}
where $\sigma_c$ can be given in km/sec. The radial distance $r$ in kpc and enclosed mass
$M(r)$ in solar masses are given (from (\ref{pars})) in terms of $x$ and $\M$
by
\begin{eqnarray}
&r/{\rm{kpc}} & = \ 0.004220\,\, \frac{\sigma_c}{\rm{km/sec}}\,\,
\left(\frac{\rm{M}_\odot/\rm{pc}^3}{\rho_c}\right)^{1/2}\,\,x,
\nonumber
\\
&M/{\rm{M}_\odot} & = \ 944.9737\,\,
\left(\frac{\sigma_c}{\rm{km/sec}}\right)^3\,\,
\left(\frac{\rm{M}_\odot/\rm{pc^3}}{\rho_c}%
\right)^{1/2}\,\,\M,
\label{physvars}\end{eqnarray}
Another important dynamical quantity is the total energy of the
stellar polytrope~\cite{TS1}:
\ba
E \ &=& \ K+U\nonumber
\\&=& \ \frac{3}{2}{\int_0}^{r}\,4\pi r^2 P(r)\,dr -{\int_0}^{r}{dr
\frac{GM(r)}{r}\frac{dM(r)}{dr}},\nonumber\\  \label{eq:E}
\ea
leading to
\be E_{poly} = -\, \frac{1}{n - 5}\Big[
\frac{3}{2}\,\frac{G\,M^2}{r}
-\Big(\frac{\rho}{\rho_c}\Big)^{1/n}\,\sigma_c\Big(\frac{3}{2}
(n + 1) M_v-(n - 2)4\,\pi\,{r}^3\,\rho_v \Big)\Big]  \label{eq:E_poly} \ee
which must be evaluated at a fixed, but arbitrary, value of $r$ marking a
cut--off scale.
\section{NFW halos}

NFW numerical simulations yield the following expression for the density
profile of virialized  galactic halo structures \cite{NFW,Mo}, \cite{LoMa}:
\begin{equation}
\rho_{_{\mathrm{NFW}}}=\frac{\delta_0\,\rho_0}{y\,\left(1+y
\right)^2},   \label{rho_NFW}
\end{equation}
where:
\begin{eqnarray} \delta_0 &&=
\frac{\Delta\,c_0^3}{3\left[\ln\,(1+c_0)-c_0/
(1+c_0)\right]},\label{delta0}\\\nonumber\\
\rho_0&&=\rho_{\mathrm{crit}}\,\Omega_0\,h^2 = 253.8 \, h^2\,\Omega_0\,
\frac{M_\odot}{\rm{kpc}^3},\label{rho0}\\
y &&= c_0\frac{r}{r_v},\label{y}
\end{eqnarray}
where $\Omega_0$ is the ratio of the total density to the critical density of the Universe, 
being one for a flat Universe.
Notice that we are using a scale parameter ($y$) that is different from that
of the stellar polytropes ($x$). The NFW virial radius, $\rvir$, is given in terms
of the virial mass, $\Mvir$, by  the condition that average halo density equals
$\Delta$ times the  cosmological density $\rho_0$:
\begin{equation}\Delta\, \rho_0 = \frac{3\,\Mvir}{4\,\pi\,\rvir^3},
\label{rvir}\end{equation}
where $\Delta$ is a model--dependent numerical factor (for a
$\Lambda$CDM model we have $\Delta\sim 100$ at $z=0$ \citep{LoHo}).
It is important to remark that this last relation between the mass and
virial radius in terms of cosmological parameters, Eq.(\ref{rvir}), is valid
for any halo model.
The concentration parameter $c_0$  can be expressed in terms of
$\Mvir$ by \citep{JZ,c0}:
\begin{equation} c_0 \approx 62.1 \times
\left(\frac{\Mvir\,h}{M_\odot}\right)^{-0.06} \label{eq:c0}
\end{equation}
hence all quantities depend on a single free parameter $\Mvir$. The mass
function and circular velocity follow from (\ref{rho_NFW}) by
$M(r)=4\pi\int{\rho\,r^2\,dr}$ and $V^2(r)=4\pi\,G\,M/r$, leading to:
\ba M =
4\,\pi\,\left(\frac{r_v}{c_0}\right)^3\,\delta_0\,\rho_0\,\left[\ln(1+y)
-\frac{y}{1+y}\right],\label{M_NFW}\\
V^2 = 4\,\pi\, G\,\delta_0\,\rho_0\,\left(\frac{r_v}{c_0}\right)^2\,\left[\frac{\ln(1+y)}{y}
-\frac{1}{1+y}\right],\label{Vsq_NFW}\ea
Given $\rho$ and $M$, the gravitational potential per unit mass and radial and
tangential pressures follow from:
\ba \Phi\,' &=& G\,\frac{M}{r^2},\label{Phi_r}\\ 
P_r\,' &=&-\rho\,\Phi\,'-\frac{2(P_r-P_\perp)}{r},\label{P_r}\ea
which combine to give:
\begin{equation}  P_r\,' =
-G\,\frac{M\,\rho}{r^2}-\frac{2\alpha(r)}{r}\,P_r,\label{P_r2}
\end{equation}
where:
\begin{equation}\alpha = \frac{P_r-P_\perp}{P_r},\label{alpha}
\end{equation}
is the anisotropy parameter $0\leq |\alpha|\leq 1$, so that
$\alpha=0$ corresponds to an isotropic velocity distribution,
since pressure is proportional to velocity dispersion. The
parameter $\alpha$ is often taken as a constant in the range
$0<\alpha< 0.2$, or given by the more realistic empirical ansatz
of Ostipov and Merritt~\cite{LoMa,OM,Merr}.

The total energy for the NFW halo follows from the general
expression (\ref{eq:E}), where $P$ should be obtained from the
integration of (\ref{P_r2}) for a specific form of $\alpha$. As
shown by \cite{LoMa,MNS}, the curves of $P=\rho\,\sigma^2$
obtained from the Ostipov--Merritt ansatz are very close to those
of the isotropic case ($\alpha=0$), hence we will consider
only isotropic NFW halos. Although analytic solutions
of (\ref{P_r2}) exist for $\alpha=0$ (see~\cite{LoMa,MNS}), we
will use instead the evaluation of (\ref{eq:E}) obtained by 
\citep{Mo} in which the leading term of the total energy is given
by:
\begin{equation}
E_{\rm{\small{NFW}}} = -\, G \, \frac{{\Mvir}^2}{2 \rvir}\, F_0
\label{eq:E_nfw}
\end{equation}
where $F_0$ has the approximate values:
\begin{equation}
F_0=\frac{2}{3}+\left(\frac{c_0}{21.5}\right)^{0.7}\end{equation}
and $c_0$ is given in terms of the virial mass by (\ref{eq:c0}).

\begin{figure}
\centering
\includegraphics[height=7cm]{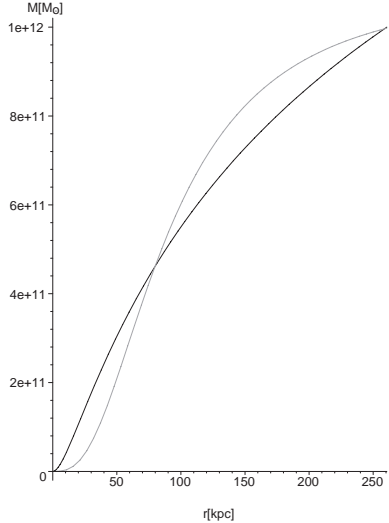}
\caption{Mass  profiles for a NFW halo with
$\Mvir=10^{12}\,M_\odot$ and $\rvir=260$ kpc (solid curve) and
compared fit stellar polytrope (dashed curve).}
\label{fig:1}       
\end{figure}

\section{Polytropic and NFW halos comparison.}
\label{sec:compara}

In order to compare stellar polytropes to NFW halos, it is
important to make various physically motivated assumptions.
First, we want both models to describe a halo of the same size,
but since the virial radius, $\rvir$, is the natural ``cut--off''
length scale at which the halo can be treated as an isolated
object in equilibrium, ``same size'' must mean same virial mass,
$\Mvir$, by equation (\ref{rvir}). 

Secondly, both models must have the same maximal value for
the rotation velocity obtained from (\ref{rotvel}) and
(\ref{Vsq_NFW}). This is a plausible assumption, as it is based
on the Tully--Fisher relation, \cite{TF}, a very well established
result that has been tested successfully for galactic systems,
showing a strong correlation between the total luminosity of a
galaxy and its maximal rotation velocity. It can be shown that the
Tully--Fisher relation has a
cosmological origin, \citep{JZ}, associated with the primordial power spectrum
of fluctuations (the so called ``cosmological  Tully-Fisher
relation''), hence it is possible to translate the correlation
between maximal rotation velocity and total luminosity to a
correlation between maximal rotation velocity and total ({\it
i.e.} virial) mass. Since, by construction we are assuming the
polytropic and NFW halos to have the same $\Mvir$, their maximal
rotation velocity must also coincide. 

Our third assumption is
that the polytropic and NFW halos, complying with the
previous requirements, also have the same total energy computed
from (\ref{eq:E_nfw}) and from (\ref{eq:E_poly}) evaluated at the
cut--off scale $r=\rvir$. The main
justification for this assumption follows from the fact that
the total energy is a fixed quantity in the collapse and subsequent
virialized equilibrium of dark matter halos~\cite{Padma1}.

Since all structural variables of the NFW halo depend only on the
virial mass, once we provide a specific value for $\Mvir$ all
variables become determined in terms of physical units by means
of the scale equation (\ref{y}). Polytropic halos, on the other
hand, lack a closed analytic expression for mass, velocity and
density profiles. In this case, equations (\ref{eq_ZMx}) (or
(\ref{eq:LE})) yield numerical solutions for these profiles
expressed in terms of the three free parameters
$\left\{\rho_c,\,\sigma_c,\,n\right\}$. The comparison of these
profiles with those of the NFW halos requires that we find
explicit values of these free parameters, so that the
conditions that we have outlined are met. Since we have selected
three comparison criteria for three parameters, we have a
mathematically consistent problem. The first constraint on
$\left\{\rho_c,\,\sigma_c,\,n\right\}$ follows by demanding that,
for a given $\Mvir$ characterizing a NFW halo, we have
$M(\rvir)=\Mvir$ for the polytropic halo obtained from the
numerical solution of (\ref{eq_ZMx}) with $M$ and $r$ given by
(\ref{physvars}). A second constraint on the polytropic
parameters follows from equating $E_{\rm{poly}}$ from
(\ref{eq:E_poly}), evaluated at $r=\rvir$ and $M=\Mvir$, with
$E_{\rm{NFW}}$ from (\ref{eq:E_nfw}). Having fixed two of the
polytropic parameters, the third one can be fixed by demanding
same maximal velocities in the curves for (\ref{rotvel}) and
(\ref{Vsq_NFW}).

\begin{table*}
\caption{Parameters characterizing the polytropes while being
compared to NFW halos}
\begin{tabular}{cccccccc}
$\log_{10}(\Mvir/M_{\odot})$ & $\rho_c\, [M_{\odot}/\rm{pc}^3]$ &
$\sigma_c\, [\rm{Km/s}]$  & $n$ & $q$ & $K_n$ & $v_{\rm{max}}\,
[\rm{Km/s}]$ & $\rvir\, [\rm{kpc}]$
\\ \hline 15 & $3.7 \times 10^{-4}$ & $982$ & $4.93$ & $1.29$ &
$4873.4$ & $1504$ & $2606.2$\\
12 & $7.5 \times 10^{-4}$ & $108$ & $4.87$ & $1.30$ &
$478.94$ & $164$ & $260.6$\\
11 & $9.0 \times 10^{-4}$ & $52$ & $4.83$ & $1.30$ &
$221.82$ & $79.1$ & $120.9$\\
10 & $1.2 \times 10^{-3}$ & $25$ & $4.82$ & $1.30$ &
$100.68$ & $38.2$ & $56.1$\\
\end{tabular}
\label{tab:t1}
\end{table*}

\begin{figure}
\centering
\includegraphics[height=7cm]{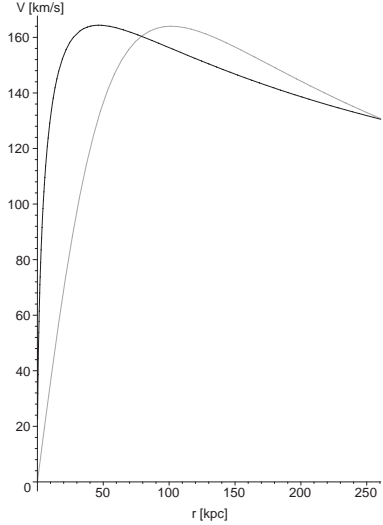}
%
\caption{Velocity profiles for the same NFW halo of figure 1 (solid line) and
its compared stellar polytrope (dashed curve).}
\label{fig:2}       
\end{figure}

Following the guidelines described above, we proceed to compare
NFW and polytropic halos for $\Mvir$ ranging from $10^{10}$ up to
$10^{15}$ solar masses. From the present comparison we find that the values for central
density, $\rho_c$, of the polytropic halos are inversely
proportional to $\Mvir$, while the values for the central
velocity dispersion, $\sigma_c$, are directly proportional to it
(this is expected, since $\sigma_c$ is a scale parameter in
self--gravitating systems). The polytropic index, $n$, is almost
constant for the selected range of $\Mvir$, showing a very small
growth as $\Mvir$ increases. This implies the same qualitative
behavior of the Tsallis parameter $q$: it is also almost constant
and is slowly increasing as the virial mass grows. It is
worthwhile mentioning that  the proportionally term $K_n$ in the
polytropic equation of state (\ref{eq:Kn}) shows a very
noticeable change, rapidly growing as $\Mvir$ increases. All
these results are displayed explicitly in table \ref{tab:t1}. The
figures depict the mass profiles,  figure \ref{fig:1}, the
velocity profiles,  figure \ref{fig:2}, and the density profiles,
figure \ref{fig:3}, for the resulting polytropes with
$\Mvir = 10^{12}\,M_\odot$, juxtaposed with the same profiles for
a NFW halo with same $\Mvir$. For other values of $\Mvir$ the
mass, velocity and density profiles are qualitatively similar to
the ones displayed in these figures.

\section{Conclusions}

Motivated by the fact that stellar polytropes are the equilibrium state in
Tsallis' non--extensive entropy formalism, we have found the structural
parameters of those stellar polytropes that allows us to compare them with NFW
halos of virial masses in the range $10^{10}<\Mvir/M_\odot<10^{15}$. The
criteria for this comparison consists in demanding that the polytropes describe a
halo having the same virial mass, virial radius, maximal rotation velocity and
total energy as the NFW halo. These three conditions are sufficient to
determine the three structural parameters
$\left\{\rho_c,\,\sigma_c,\,n\right\}$ of the polytropic model; the results are 
displayed in Table \ref{tab:t1}.

\begin{figure}
\centering
\includegraphics[height=7cm]{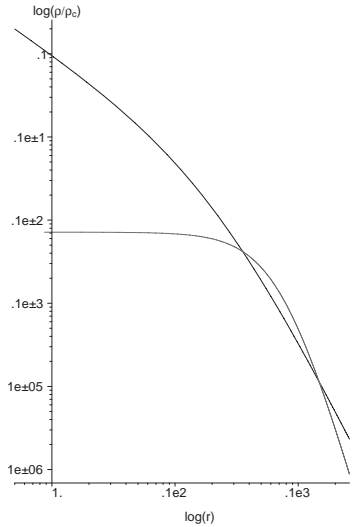}
\caption{Density profiles for the same NFW halo of figures 1 and
2 (solid curve) and its compared stellar polytrope (dashed curve).}
\label{fig:3}       
\end{figure}

It is important to emphasize that the criteria that determine these
polytropes are based on physically motivated
assumptions: the virial radius and mass are the natural
parameters characterizing the size of a given halo, same maximal
velocity follows from the Tully--Fisher relation, while same total
energy follows from the virilization process. As shown in
Figure \ref{fig:1}, the mass distribution of the polytrope grows
much slower than that of the NFW halo up to a large radius (100
kpc) containing the core and the region where visible matter
concentrates. Hence, as shown by Figure \ref{fig:2}, the velocity
profile of the polytrope is much less steep in the same region
than that of the NFW halo. These features are consistent with the
fact that NFW profiles predict more dark matter mass
concentration than what is actually observed in a large sample of
galaxies~\citep{vera,LSB,JZ}. Also, as shown in Figure
\ref{fig:3}, the obtained polytropes have flat cores, very
similar to the flat isothermal cores observed in LSB galaxies (as
a contrast, the cuspy cores of NFW halos seem to be at odds with
these observations~\citep{cdm_problems_1, cdm_problems_2,
cdm_problems_3}, also \citep{vera, LSB}). This flat density
core is a nice property, which combined with reasonable mass and
velocity profiles, qualifies these polytropes as reasonable
(albeit idealized) models of halo structures.

However, in spite of their nice theoretical properties ({\it
i.e.} their connection to Tsallis' formalism) and reasonable
similarity with equivalent NFW halos, the stellar polytropes we have
examined are very idealized configurations and so we are not
claiming that they provide a realistic description of halo
structures. Instead, we suggest that their described features and their
connection with Tsallis' formalism might
indicate that the latter could yield useful information in
understanding the evolution and virialization process of dark
matter. Although it is necessary to pursue this idea by
means of more sophisticated methods, including the use of
numerical simulations along the lines pioneered by \citep{TS2},
the simple approach we have presented has already given
interesting results. For example, with respect to the parameter, $q$, we recall that it is
a free parameter of the Tsallis' non-extensive thermodynamics and
which has not been fixed for the cosmological case. In this work,
by using such statistic in cosmological systems, we are able to
determine its behavior as a function of the virial mass, and turns
out to be almost constant, with a values around $q\approx 1.2$.
This result could be used in other contexts where the extended
statistic is also applied \cite{Tsallis2}.

It is well known~\citep{B-T} that stellar polytropes with
$n<5$ (like King halos) have a finite cut--off scale and finite
total mass, though for the polytropes that we have
studied this cut--off scale is much larger than $\rvir$ (just as
the ``tidal radius'' of King halos is much larger than their
virial radius). However, as shown in~\citep{TS1,TS2}, polytropes
characterized by this polytropic index correspond to stable
equilibrium states that are generically free from undergoing
gravothermal instability. 

As mentioned, the results presented in the present work show that
a dark matter halo made out of matter which satisfies a polytropic like 
equation of state, describes the halo in a way as good as the description
obtained from the NFW numerical simmulations, that is their paradigm. Furthermore,
our description is even nicer as long as it does not have great density growths 
near the center. However, these results does not directly imply that the dark matter halo
do obey a non-extensive entropy formalism. Further tests and expriments are needed
in order to consider that such formalism is the one describing the thermodynamics
of actual dark matter halos. At the moment, this idea is a possibility which
is reinforced by our analysis. We believe these properties to be very
encouraging and are currently engaged in a more detailed
examination of these polytropes~\citep{sigue}.


\acknowledgements It is a pleasure to participate with the present work
in the Festschrift in honor of our colleague Mike Ryan. This work was partly supported by CONACyT
M\'exico, under grants 32138-E and 34407-E. We also acknowledge
support from grants DGAPA-UNAM IN-122002, IN117803, and IN109001.
JZ acknowledges support from DGEP-UNAM and CONACyT scholarships.



\begin{thebibliography}{}


\bibitem{Blok1}de Blok, W. J. G., MacGaugh, S. S., Bosma, A., and
Rubin, V. C., 2001, ApJ, 552, L23; de Blok, W. J. G., MacGaugh, S.
S., and Rubin, V. C., 2001, ApJ, 122, 2396; Binney,  J. J., and
Evans, N. W., 2001, MNRAS, 327, L27; Blais-Ouellette, Carignan,
C., and Amram, P.,  2002, E-print astro-ph/0203146; Borriello, A.
and Salucci, P. MNRAS, 323, 285; Borriello, Salucci,  Danese 2002,
MNRAS 341, 1109; P. and Burkert, A. 2000, ApJ, 537, L9; P. 2001,
MNRAS, 320, L1; P., Walter, F., and Borriello, A., 2002 E-print
astro-ph/0206304.

\bibitem{self}Spergel, D. N., Steinhardt, P. J.,
2000, Phys. Rev. Lett., 84, 3760,
E-Print: astro-ph/9909386


\bibitem{warm}Colin, P., Avila-Reese, V., and Valenzuela, O.,
2001,ApJ, 542, 622, E-print: astro-ph/0004115

\bibitem{DMCQG}Guzm\'{a}n, F. S., and Matos, T., 2000, Class. Quantum Grav. {\bf 17},
L9; Guzm\'{a}n, F. S., and Ure\~{n}a-L\'{o}pez, L. A., 2003, Phys.
Rev. D, 68, 024023; Matos, T., and Guzm\'{a}n, F. S., 2000, Ann.
Phys. (Leipzig), 9, SI-133; Matos, T., Guzm\'{a}n, F. S., and
N\'{u}\~{n}ez, D., 2000, Phys. Rev. D, 62, 061301; Matos, T., and
Guzm\'{a}n, F. S., 2001, Class. Quantum Grav., 18, 5055; Matos,
T., and Ure\~{n}a-L\'{o}pez, L. A., Class. Quantum Grav., 2000,
17, L75; Matos, T., and Ure\~{n}a-L\'{o}pez, L. A., 2001, Phys.
Rev. D, 63, 63506; Matos, T., and Ure\~{n}a-L\'{o}pez, L. A.,
2002, Phys. Lett. B 538, 246; Guzm\'{a}n, F. S., and
Ure\~{n}a-L\'{o}pez, L. A., 2004, Phys. Rev. D, 69, 124033;
Guzm\'{a}n, F. S., 2004, Phys. Rev. D, 69, in press;

\bibitem{Ruffini}Ruffini, R., and Bonazzola S., 1969, Phys. Rev. D, 187, 1767;

\bibitem{Padma2}Padmanabhan, T. 1990, Phys. Rep. 188, 285
\bibitem{Padma3}Padmanabhan, T. 2000, Theoretical Astrophysics, Volume I:
Astrophysical Processes, ed. Cambridge University Press

\bibitem{Tsallis}Tsallis, C. 1999, Braz J Phys, 29, 1

\bibitem{PL}Plastino, A. R., \& and A Plastino, A. 1993, Phys Lett A, 174, 384

\bibitem{TS1}Taruya, A., \& Sakagami, M. 2002, Physica, A 307, 185, e-Print:cond-mat/0107494;
\bibitem{TS2}--, 2003a, Phys. Rev. Lett., 90, 181101; See also --, 2003, e-Print:cond-mat/0310082

\bibitem{B-T}Binney, J., \&  S. Tremaine, S. 1987, Galactic dynamics, ed.
Princeton Univ.

\bibitem{NFW}Navarro, J. F., \& Frenk, C. S., \& White, S. D. M. 1997,
ApJ, 490, 493, e-Print:astro-ph/9611107

\bibitem{Mo}Mo, H. J., \& Mao, S., \& White, S. D. M. 1998, MNRAS,
295, 319, e-Print:astro-ph/9707093.

\bibitem{LoMa}Lokas, E. L., \& Mamon, G. 2001, MNRAS, 321,  155

\bibitem{LoHo}Lokas, E. L., \& Hoffman, Y. 2001, e-Print:astro-ph/0108283. See also
Lokas, E. L. 2001, Acta Phys.Polon.,  B32, 3643

\bibitem{JZ}Avila--Reese, V.,  et al, 2003, A\&A, 412,
633,  e-Print:astro-ph/0305516.

\bibitem{c0}Eke, V. R., \& Navarro, J. F., \& and Steinmetz, M. 2001, Ap J, 554, 114

\bibitem{OM}Ostipov, L. P. 1979, PAZh, 5, 77

\bibitem{Merr}Merritt, D. 1985, ApJ, 90, 1027

\bibitem{MNS}Matos, T., \& N\'u\~nez, D.,\& Sussman, R. A. 2004, 
Gen. Rel. Grav., in press, e-Print:astro-ph/0402157

\bibitem{TF}Tully, R. B., \& Fisher, J. R. 1977, A\&A, 54, 661

\bibitem{Padma1}Padmanabhan, T. 1993, Structure formation in the universe,
Cambridge University Press

\bibitem{vera}Rubin, V. C., \& de Blok, W. J. G., 2001, ApJ, 122, 2381

\bibitem{LSB}Binney,  J. J., \& Evans, N. W. 2001, MNRAS, 327, L27

\bibitem{cdm_problems_1}  Moore, B. 1994, Nature, 370, 629

\bibitem{cdm_problems_2}Flores, R., \& Primack, J. P. 1994, ApJ, 427, L1

\bibitem{cdm_problems_3}Burkert, A. 1997, Aspects of Dark  Matter in Astro-and
Particle Physics (ed. H.V. Klapdor-Kleingrothaus, H. V., \&
Ramachers, Y.), e-Print: astro-ph/9703057.

\bibitem[Tsallis 2001]{Tsallis2}-- 2001, ( Eds. Abe, S., \&
Okamoto, Y.  Nonextensive Statistical Mechanics and its
Applications, Springer, Berlin, 2001)

\bibitem{sigue}Cabral-Rosetti, L. G., et. al., in preparation.


\end{thebibliography}
\end{document}